# Theoretical Study of Physisorption of Nucleobases on Boron Nitride Nanotubes: A New Class of Hybrid Nano-Bio Materials

**Saikat Mukhopadhyay[1], S. Gowtham[1], Ralph H. Scheicher[2], Ravindra Pandey[1]* and Shashi P. Karna[3]**

[1]*Department of Physics, Michigan Technological University. Houghton, MI 49931, USA.*

[2]*Department of Physics and Materials Science, Uppsala University, SE-751 21 Uppsala, Sweden.*

[3]*US Army Research Laboratory, Weapons and Materials Research Directorate, ATTN: RDRL-WM, Aberdeen Proving Ground, MD 21005-5069, U.S.A.*

March 8, 2010

*To whom correspondence should be addressed: E-mail:* pandey@mtu.edu

**Accepted in *Nanotechnology (March 01, 2010)***

**Abstract**

We investigate the adsorption of the nucleic acid bases - adenine (A), guanine (G), cytosine (C), thymine (T) and uracil (U) - on the outer wall of a high curvature semiconducting single-walled boron nitride nanotube (BNNT) by first-principles density functional theory calculations. The calculated binding energy shows the order: G>A≈C≈T≈U implying that the interaction strength of the high curvature BNNT with the nucleobases, G being an exception, is nearly the same. A higher binding energy for the G-BNNT conjugate appears to result from a hybridization of the molecular orbitals of G and BNNT. A smaller energy gap predicted for the G-BNNT conjugate relative to that of the pristine BNNT may be useful in application of this class of biofunctional materials to the design of the next generation sensing devices.

## 1. Introduction:

Boron nitride nanotubes (BNNTs) have been the focus of several experimental and theoretical studies [e.g. References 1-3] due to their potential applications in high speed electronics. BNNTs are a typical member of III-V compound semiconductors with morphology similar to that of carbon nanotubes (CNTs) but with their own distinct properties. A tubular structure of BN can be formed by rolling up a sheet of hexagonal rings, with boron and nitrogen in equal proportions possessing peculiar electrical [4], optical [5] and thermal [6] properties, which drastically differ from those of CNTs.

The nucleic acid bases, on the other hand, being the key components of the genetic macromolecules - deoxyribonucleic acid (DNA) and ribonucleic acid (RNA) - play a central role in all biological systems and thus have been a focus of intense research activities over the past five decades. Recently, there has been a keen interest in understanding the interaction between nucleobases and matter, especially nanostructured materials such as carbon nanotubes [7-16] due to the potential application of the unique signature of the latter in probing the structural and conformational changes [17, 18] of the former, and hence leading to new detection mechanism [19] and medical diagnostic tools.

Very recently, a thiol-modified DNA was used to obtain high concentration BNNT aqueous solutions assuming the interaction between DNA and multi-walled BNNT to be strong [20]. Analysis of the transmission electron microscopy measurements showed that the thiol-modified DNA wraps around the tubular surface of BN. The tubular surface of BN consists of dissimilar atoms and, thus, its interaction with the nucleobases may show different characteristics as compared to that observed in case of either graphene or CNTs.

Previously, the interaction of nucleobases with graphene [7, 11] and CNTs [12] was predicted to be dominated by van der Waals (vdW) forces as the binding energy is seen to increase with the polarizability of the nucleobases. The charge transfer between the nucleobases and CNTs was found to be negligible. In the present study, our motivation is to systematically investigate the self-organization of the nucleobases onto the tubular surface of BN and identify factors playing a role on the differences in the interaction for different base molecules. Wherever possible, we compare the results of our study with the previous studies on CNTs.

## 2. Methodology:

We consider a high curvature (5, 0) single-walled BNNT of diameter of 0.416 nm, which has been predicted to be stable by theoretical calculations [21]. All calculations were performed by employing the plane wave pseudopotential approach within the local density approximation (LDA) [22] of density functional theory (DFT) [23, 24]. The Vienna ab initio Simulation Package (VASP) was used [25, 26] with the energy cut off of 850 eV and 0.03 eV/Å for its gradient. The periodically repeated BNNT units were separated by 15 Å of vacuum to avoid interaction between them. The (1x1x3) Monkhorst Pack grid [27] was used for *k*-point sampling of the Brillouin zone. The average B-N bond length in the optimized configuration of the pristine BNNT is 1.44 Å, consistent with previously reported DFT calculations [31 and references therein].

In the calculations of the energy surface describing the interaction of the nucleobases with BNNT, the nucleobases were allowed to approach the tubular surface in the direction perpendicular to the axis of the tube. In order to simulate an electronic environment resembling more closely the situation in DNA and RNA, the N atom of the base molecules linked to the sugar ring in nucleic acid was terminated with a methyl group. There is an additional benefit of introducing the small magnitude of steric hindrance due to the attached methyl group. It will help us to imitate a more probable situation in which a nucleobase in a strand would interact with the surface of the BNNT rather than an isolated nucleobase interacting with BNNT. For simulations based on force fields [9-14], it is certainly possible to include all constituents of DNA, including the sugar-phosphate backbone. In the present first-principles study however, simulation of the nucleobases attached to the backbone (i.e. sugar + phosphate group) is computationally rather expensive.

The optimized configurations of the nucleobases-BNNT conjugate systems were obtained following a similar scheme as employed in the previous study of BNNT-CNT complex [12]. It consisted of (i) an initial force relaxation calculation step to determine the preferred orientation and optimum height of the planar base molecule relative to the surface of the BNNT (ii) calculations of the potential energy surface [Figure 1] for nucleobase-BNNT interaction by translating the relaxed base molecules parallel to the BNNT surface covering a surface area 4.26 Å in height, $70^0$ in width, and containing a mesh of 230 scan points. The separation between base molecule and the surface of the BNNT was held fixed at the optimum height determined in step (i). (iii) a $360^0$ rotation of the base molecules in steps of $5^0$ to probe the energy dependence on the orientation of the base molecules with respect to the underlying BNNT surface; and (iv) a full optimization of the conjugate system in which all atoms were free to relax.

Certainly, in a potential energy surface scan, some lateral restriction against sliding must be applied. However, it is true that, in principle, reorientation through rotations should be considered. This was not done here, for the following two reasons: first, regarding rotations around any axis that lies in the plane of the nucleobase (comparable to "roll" and "pitch" for airplanes), the preferred orientation of the nucleobase plane relative to the tube surface is parallel in order to maximize the attractive van der Waals interaction while minimizing the repulsive interaction from overlapping electron clouds. Second, regarding rotations around the axis that goes perpendicular through the plane of the nucleobase (comparable to "yaw" for airplanes), at least for the larger purine base molecules, the preferred orientation is such that their longer dimension is aligned with the tube axis (as seen in the equilibrium geometries shown in Figure 2) again in order to maximize the attractive interaction with the tube surface.

It should be pointed that LDA due to a lack of the description of dispersive forces is, in principle, not the most optimal choice for calculating interaction energies of systems governed by vdW forces. However, more sophisticated methods, such as many-body perturbation theory, which are more suitable for describing long range forces, become prohibitively expensive for complex systems as considered here. Earlier studies [28, 29] have shown that, unlike the generalized gradient approximation (GGA) [30] for which the binding for vdW bound systems does not exist, the LDA approximation does indeed provide reasonably good description of the dispersive interactions. Also a recent study [7] on the adsorption of adenine on graphite suggests that the potential energy surface obtained by using LDA and GGA with a modified version of the London dispersion formula for vdW interactions is effectively indistinguishable. Additionally, the LDA equilibrium distance between adenine and graphene obtained by LDA is found to be equal to that obtained using GGA + vdW level of theory. This gives us confidence in the results obtained in the present study to be reasonably accurate in describing nucleobase-BNNT interaction.

**3. Results and Discussion:**

The calculated base-BNNT binding energy, $E_b$, the equilibrium base-BNNT distance, and the band gap of the corresponding base-BNNT complex are listed in Table 1. The optimized configuration of the nucleobase-BNNT conjugates are shown in Figure 2. It should be noted that the base molecule was allowed to approach the tubular surface along the axes perpendicular to that of the tube while obtaining the potential energy surface. In addition to that as mentioned in the methodology section, we have scanned the surface of the tube [step (ii)] shown in Figure 1. After that we have rotated the base molecule to check if any particular orientation of the base molecule is preferred [optimization step (iii)] and at the end the whole system was optimized relaxing BNNT and the nucleobases [optimization step (iv)]. The equilibrium configurations shown in Figure 2 were the energetically most favorable one.

None of the nucleobases show a perfect Bernal's AB stacking. This feature matches with what was found in the interaction of the nucleobases and SWCNT. There is, however, a slight difference in the stacking of the nucleobases between BNNT and SWCNT, because BNNT possess a hetero-nucleic surface unlike SWCNT. The partially negatively charged oxygen atom in guanine can interact electrostatically with the polar network of this heteronucleic BNNT surface, specifically in an attractive manner with the partial positive charges on boron, and repulsively with the partial negative charges on nitrogen. This could help to explain our theoretical observation that G+BNNT differs in several ways from A+BNNT, since in the latter combination, adenine lacks the oxygen atom. Thus, the deviation in the stacking arrangement, the higher binding energy (see Table 1), and the slight tilting angle could all be consequences of that interaction between the oxygen atom of guanine and the polar network of the BNNT.

The nearest-neighbor distance ($R_{base\text{-}BNNT}$) of the individual atoms of the nucleobases from the tubular surface atoms is plotted in Figure 3 which is found to depend on the nucleobases. We note that $R_{base\text{-}BNNT}$ is comparable to the average distance of organic molecules including amino functional groups and 2, 4, 6-trinitrotoluene physisorbed on BNNTs [31, 32].

Figure 4 shows the energy surface representing the interaction of nucleobases with BNNT. Here, the distance is taken to be the separation from the center of mass of the tubular configuration to the center of mass of the nucleobases. The asymptotic limit of the energy surface is used to calculate the binding energy ($E_b$) of the system (Table 1) in which the base molecule is moved away from BNNT along the direction perpendicular to the tubular axis. The binding energy data is presented in Table 1.

***Table 1: Binding energy ($E_b$), band gap, and nearest-neighbor distance ($R_{base\text{-}BNNT}$) of nucleobase conjugated BNNT. The calculated LDA band gap of the pristine (5, 0) BNNT is 2.2 eV.***

| *System* | $E_b$ *(eV)* | $R_{base\text{-}BNNT}$ *(Å)* | *Band gap (eV)* |
|---|---|---|---|
| *G+BNNT* | *0.42* | *2.49* | *1.0* |
| *A+BNNT* | *0.32* | *3.06* | *1.7* |
| *C+BNNT* | *0.31* | *2.96* | *1.8* |
| *T+BNNT* | *0.29* | *2.55* | *2.0* |
| *U+BNNT* | *0.29* | *2.86* | *2.1* |

The magnitude of the calculated binding energy exhibit the following order: G>A≈C≈T≈U. It is worth noting that the binding energy of the nucleobases interacting with a high curvature CNT [12] followed the order of G>A>T>C≈U. Since $E_b$ associated with CNTs was found to be correlated with the polarizability of individual bases, it was suggested that the interaction of nucleobases with CNTs was governed by the dispersive force like vdW which varies with the polarizability of the interacting entity. The calculated polarizability values of G, A, C, T, U are 131.2, 123.7, 111.4, 108.5 and 97.6 $e^2 a_0^2 E_h^{-1}$, respectively, at the Hartree-Fock level of theory together with the second order Møller–Plesset corrections [12].

Figure 5 shows the total charge density plot of the representative conjugate system of G physisorbed on BNNT. The Bader charge analysis does not show a noticeable charge transfer in the conjugate system relative to the pristine BNNT and individual nucleobases; change in the total charge of the nucleobases being quite small ($< 10^{-2}$ *e*). This is in contrast to the cases of covalent functionalized

BNNTs [32-34] where a significant charge transfer of the order of 0.36 *e* from the organic molecule such as $NH_3$ and amino functional groups to BNNT was reported. Our results are consistent with the case of 2, 4, 6-trinitrotoluene physisorbed on BNNTs reporting a very small charge transfer in the system [31].

In order to further understand the underlying interaction between the nucleobases and BNNT, we also calculated the polarizability of a BN sheet which comes out to be 265.7 $e^2a_0^2$ $E_h^{-1}$ at the LDA level of theory. The polarizability of a BN sheet is therefore significantly smaller than 402.2 $e^2a_0^2$ $E_h^{-1}$ calculated for graphene at the same level of theory. This suggests that the tubular surface of a BNNT can be expected to be less polarizable than that of a CNT which, in turn, would lead to relatively weaker vdW interactions between BNNT and nucleobases. This is reflected in the calculated binding energy values of physisorbed nucleobases on BNNT which are lower in magnitude as compared to those associated with CNTs. For example, the calculated binding energy of G+BNNT conjugate is 0.4 eV while the corresponding value for the G+CNT conjugate is 0.5 eV.

A comparison of the present results with those from a previous study, it is found that the binding energy of nucleobases with (7, 7) BNNT is significantly higher than that with CNTs [34]. This LDA study using numerical atomic orbitals reported a binding energy of about 1 eV for the G+BNNT conjugate system. This clearly suggests that the lower surface curvature of the (7, 7) BNNT (with a diameter of 9.60 Å) leads to a stronger interaction with the nucleobases than a large surface curvature for the (5, 0) BNNT (with a diameter of 4.16 Å) considered in the present study. A similar trend in the effect of the surface curvature on the binding energy between nucleobase and carbon nanostructures, graphene [11] and CNT [12] was noted in previous studies.

The semiconducting nature of BNNT with a band gap of about 2.2 eV can be seen in the calculated density of states shown in Figure 6. This is in agreement with the recent LDA calculations on the pristine (4, 0) BNNTs reporting a direct band gap of about 2.0 eV [36]. Both the top of the valence edge and bottom of the conduction edge of BNNT are associated with the N p-orbitals. The asymmetry in DOS appears to be due to the difference in coupling strength between the π-orbitals of BNNT and the base molecule in the valance band compared to the conduction band. In the former case contribution from the nucleobases dominates and contributions from the BNNT dominate in the latter case. The appearance of the mid gap states (Figure 6) in the conjugated BNNT represents a mixing of electronic states of the nucleobases and BNNT separated at about at about 2.5Å. It may be noted that the covalent interaction at the separation of 2.5Å will be very weak [37]. On the other hand a very small charge transfer from BNNT to oxygen of guanine may indicate the presence of a relatively weaker electrostatic interaction between BNNT and guanine. The interaction between BNNT and nucleobases is however essentially dominated by the vdW forces.

## 4. Summary and Conclusion.

In summary, we have investigated the interaction of the nucleobases on a high curvature, zigzag (5,0) BNNT by first-principles DFT method. Our calculations show that, except G, the base molecules (A, C, T, U) of DNA and RNA exhibit almost similar interaction strengths when physisorbed on BNNT. It is also observed that the binding energy of the base molecules not only depends upon their individual polarizability but also marginally depends on the degrees of mixing of electronic states with the tubular surface of BNNT. The strong binding of the BNNT with G compared to the other nucleobases suggests that this interaction can be used in sensing and also for distinguishing this base molecule from other nucleic acid bases.

**Acknowledgments**

Helpful discussions with Haiying He are thankfully acknowledged. The work at Michigan Technological University was performed under support by Army Research Office through contract number W911NF-09-1-0221. RHS acknowledges financial support from Wenner-Gren Foundations in Stockholm.

**References:**

[1] A. Rubio, J. L. Corkill, M. L. Cohen "*Theory of graphitic boron nanotube*" Phys. Rev. B 49, 5081, 1994.

[2] N. G. Chopra, R. J. Luyken , K. Cherrey , V. H. Crespi , M. L. Cohen , S. G. Louie , and A. Zettl "*Boron Nitride Nanotubes*" Science 269, 966, 1995.

[3] S. Riikonen, A. S. Foster, A. V. Krasheninnikov, R. M. Nieminen *"Computational study of boron nitride nanotube synthesis: How catalyst morphology stabilizes the boron nitride bond"* Phys. Rev. B 80, 155429, 2009.

[4] R. Pati, P. Panigrahi, P. P. Pal, B. Akdim, R. Pachter "*Gate field induced electronic current modulation in a single wall boron nitride nanotube: Molecular scale field effect transistor"* Chem. Phys. Lett. 482, 312 , 2009.

[5] J. S. Lauret, R. Arenal, F. Ducastelle, A. Loiseau, M. Cau, B. Attal-Tretout, E. Rosencher, L. Goux-Capes *"Optical Transitions in Single-Wall Boron Nitride Nanotubes"* Phys. Rev. Lett. 94, 037405,2005.

[6] Y. Xiao, X. H. Yan, J. Xiang, Y. L. Mao, Y. Zhang, J. X. Cao, J. W. Ding "*Specific heat of single-walled boron nitride nanotubes"* Appl. Phys. Lett. 84, 4626, 2004.

[7] F. Ortmann, W. G. Schmidt, F. Bechstedt "*Attracted by Long-Range Electron Correlation: Adenine on Graphite*" Phy. Rev. Lett 95, 186101, 2005.

[8] E. S. Jeng, A. E. Moll, A. C. Roy, J. B. Gastala, M. S. Strano "*Detection of DNA Hybridization Using the Near-Infrared Band-Gap Fluorescence of Single-Walled Carbon Nanotubes"* Nano Lett. *6, 371, 2006.*

[9] S. Meng, P. Maragakis, C. Papaloukas, E. Kaxiras "*DNA Nucleoside Interaction and Identification with Carbon Nanotubes"* Nano Lett. 7, 45, 2007.

[10] A. N. Enyashin, S. Gemming, G. Seifert "*DNA-wrapped carbon nanotubes*" Nanotechnology 18, 245702, 2007.

[11] S. Gowtham, R. H. Scheicher, R. Ahuja, R. Pandey, S. P. Karna "*Physisorption of nucleobases on graphene: Density-functional calculations"* Phys. Rev. B 76, 033401, 2007.

[12] S. Gowtham, R. H. Scheicher, Ravindra Pandey, S. P. Karna, R. Ahuja "*First-principles study of physisorption of nucleic acid bases on small-diameter carbon nanotubes"* Nanotechnology 19, 125701 , 2008.

[13] R. R. Johnson, A. T. C. Johnson, M. L. Klein "*Probing the Structure of DNA-Carbon Nanotube Hybrids with Molecular Dynamics*" Nano Lett. 8, 69, 2008.

[14] *A. Das, A. K. Sood, P. K. Maiti, M. Das, R. Varadarajan, C. N. R. Rao "Binding of nucleobases with single-walled carbon nanotubes: Theory and experiment"* Chem. Phys. Lett. *453,* 266, 2008.

[15] R. R. Johnson, A. Kohlmeyer, A. T. C. Johnson, M. L. Klein "*Free Energy Landscape of a DNA-Carbon Nanotube Hybrid Using Replica Exchange Molecular Dynamics*" Nano Lett. 9, 537, 2009.

[16] X. M. Tu, S. Manohar, A. Jagota, M. Zheng "*DNA sequence motifs for structure-specific recognition and separation of carbon nanotubes*" Nature 460, 250, 2009.

[17] V. G. Vaidyanathan, B. U. Nair "*Synthesis, characterization and binding studies of chromium(III) complex containing an intercalating ligand with DNA"* J. Inorg. Biochemistry 95, 334, 2003.

[18] A. Rajendran , C. J. Magesh, P. T. Perumal "*DNA–DNA cross-linking mediated by bifunctional [SalenAlIII]+ complex"* Biochimica et Biophysica Acta, 1780, 282,2008.

[19] D. A. Heller, E. S. Jeng, T. K. Yeung, B. M. Martinez, A. E. Moll, J. B. Gastala, M. S. Strano "*Optical Detection of DNA Conformational Polymorphism on Single-Walled Carbon Nanotubes*" Science 311, 508, 2006.

[20] C. Zhi, Y. Bando, W. Wang, C. Tang, H. Kuwahara, D. Golberg "*DNA-Mediated Assembly of Boron Nitride Nanotubes"* Chem. Asian J. 2, 1581 (2007).

[21] H. J. Xiang, J. Yang, J. G. Hou, Q. Zhu "*First-principles study of small-radius single-walled BN nanotubes*" Phys Rev B 68, 035427, 2003.

[22] J. P. Perdew, A. Zunger " *Self-interaction correction to density-functional approximations for many-electron systems*" Phys. Rev B 23, 5048, 1981.

[23] P. Hohenberg, W. Kohn "Inhomogeneous *Electron Gas*" Phys. Rev. 136, B865, 1964.

[24] W. Kohn, L. J. Sham " *Self-Consistent Equations Including Exchange and Correlation Effects*" Phys. Rev. 140, A1133, 1965.

[25] G. Kresse, J. Furthmüller "*Efficiency of ab-initio total energy calculations for metals and semiconductors using a plane-wave basis set"* Comput Mat Sci 6 , 15 1996.

[26] G. Kresse, D. Joubert *"From ultrasoft pseudopotentials to the projector augmented-wave method"Phys. Rev. B 59, 1758, 1999.*

[27] H. J. Monkhorst, J. D. Pack " *Special points for Brillouin-zone integrations*" Phys. Rev. B 13, 5188, 1976.

[28] M. Simeoni, C. D. Luca, S. Picozzi, S. Santucci, B, Delley "*Interaction between zigzag single-wall carbon nanotubes and polymers: A density-functional study"* J. Chem. Phys. 122, 214710, 2005.

[29] F. Tournus, S. Latil, M. I. Heggie, J. C. Charlier *"π- stacking interaction between carbon nanotubes and organic molecules"* Phys. Rev. B 72, 075431, 2005.

[30] J. P. Perdew, J.A, Chevary, S. H. Vosko, K. A. Jackson, M. R. Pederson, D. J. Sing, C. Fiolhais "*Atoms, molecules, solids, and surfaces: Applications of the generalized gradient approximation for exchange and correlation*" Phys. Rev. B 46, 6671,1992.

[31] B. Akdim, S. N. Kim, R. R. Naik, B. Maruyama, M. J. Pender, R Pachter "*Understanding effects of molecular adsorption at a single-wall boron nitride nanotube interface from density functional theory calculations"* Nanotechnology 20, 355705, 2009.

[32] X. Wu, W. An, X. C. Zeng "*Chemical Functionalization of Boron-Nitride Nanotubes with NH3 and Amino Functional Group* " J. Am. Chem. Soc. 128, 12001.2006.

[33] C. Zhi, Y. Bando, C. Tang, S. Honda, K. Sato, H. Kuwahara, D. Golberg "*Covalent Functionalization: Towards Soluble Multiwalled Boron Nitride Nanotubes"* Angew. Chem., Int. Ed. 44, 7932, 2005.

[34] J. Zheng, W. Song, L. Wang, J. Lu, G. Luo, J. Zhou, R. Qin, H. Li, Z. Gao, L. Lai, G. Li, W. N. Mei "*Adsorption of Nucleic Acid Bases and Amino Acids on Single-Walled Carbon and Boron Nitride Nanotubes: A First-Principles Study* " J. Nanoscience and Nanotechnology 9, 6376, 2009.

[35] C. Zhi, Y. Bando, C. Tang, D. Golberg "*Engineering of electronic structure of boron-nitride nanotubes by covalent functionalization*" Phys. Rev. B 74, 153413 (2006).

[36] Z. Zhang, W. Guo, Y. Dai "*Stability and electronic properties of small boron nitride nanotubes*." J. Appl. Phys 105, 84312 ,2009.

[37] S. J. Grabowski, W. A. Sokalski, E. Dyguda, J. Leszczyński "*Quantitative Classification of Covalent and Noncovalent H-Bonds"* J. Phys. Chem. B 110, 6444, 2006.

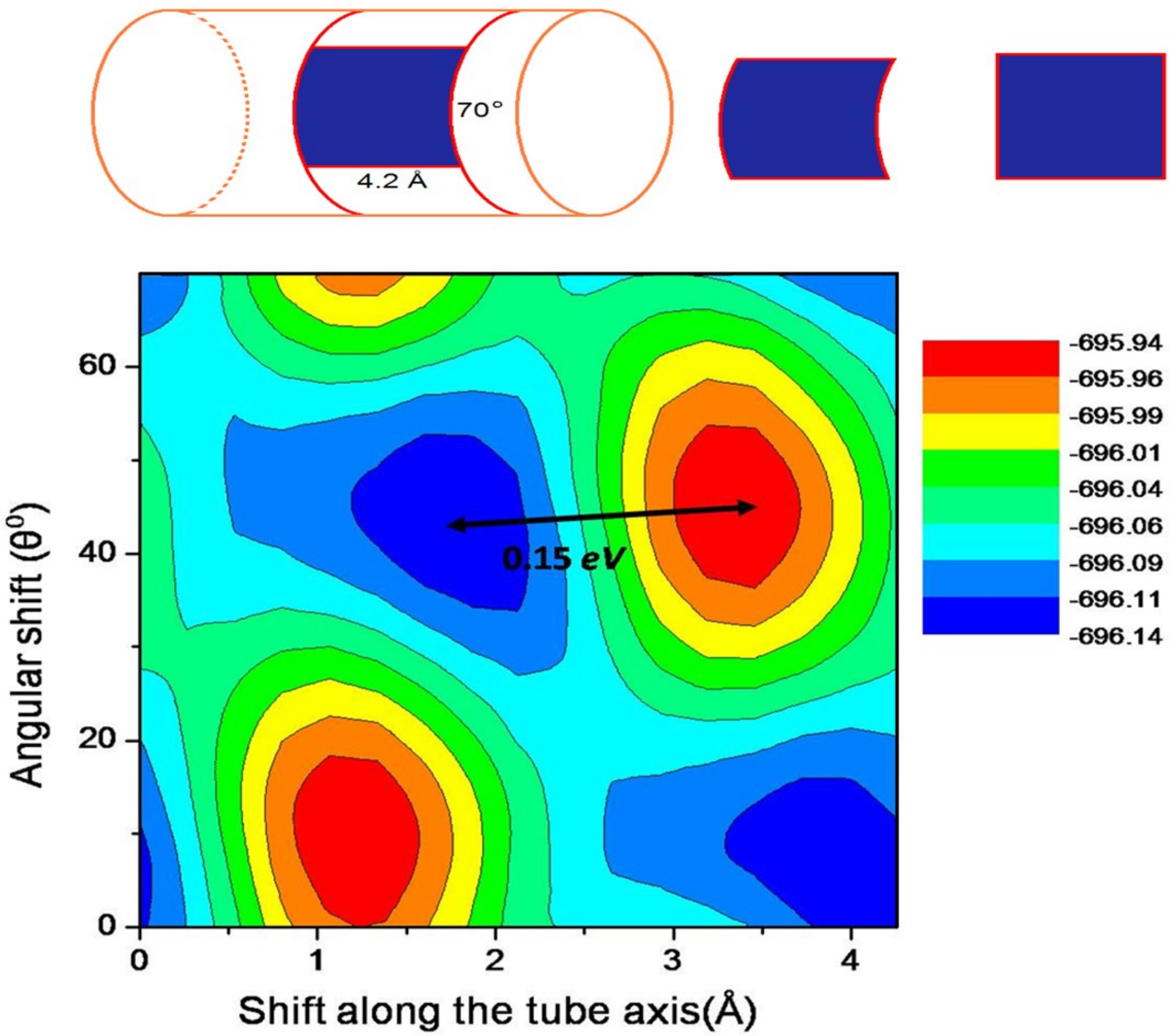

***Figure 1: Potential energy surface plot (eV) for guanine scanning the surface of BNNT. The specifications of the Scanned area (rectangle highlighted by dark blue) are shown above. The energy barrier between two adjacent global minima is 0.15 eV as indicated by the arrows. Qualitatively similar features were found in case of the other four nucleobases.***

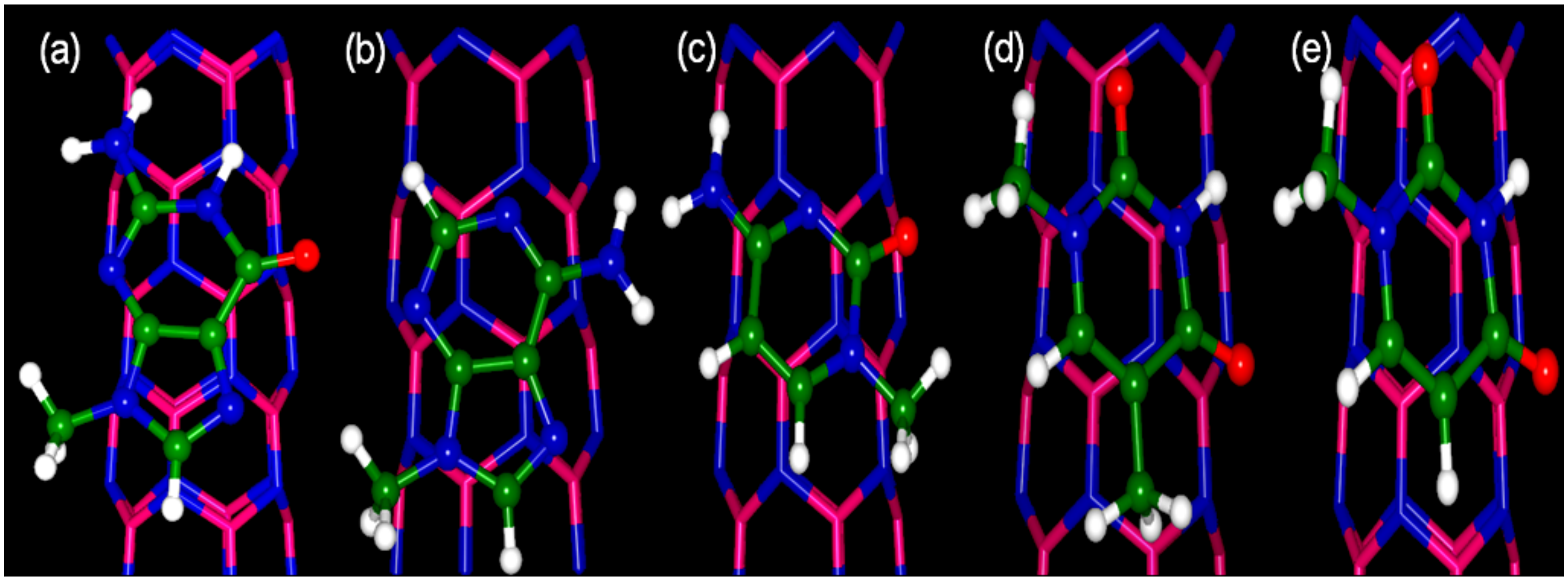

***Figure 2: Equilibrium geometry of physisorbed nucleobases on the surface BNNT. (a) guanine, (b) adenine, (c) cytosine, (d) thymine and (e) uracil. The pink, blue, green, red and white colors represent B, N, C, O and H, respectively.***

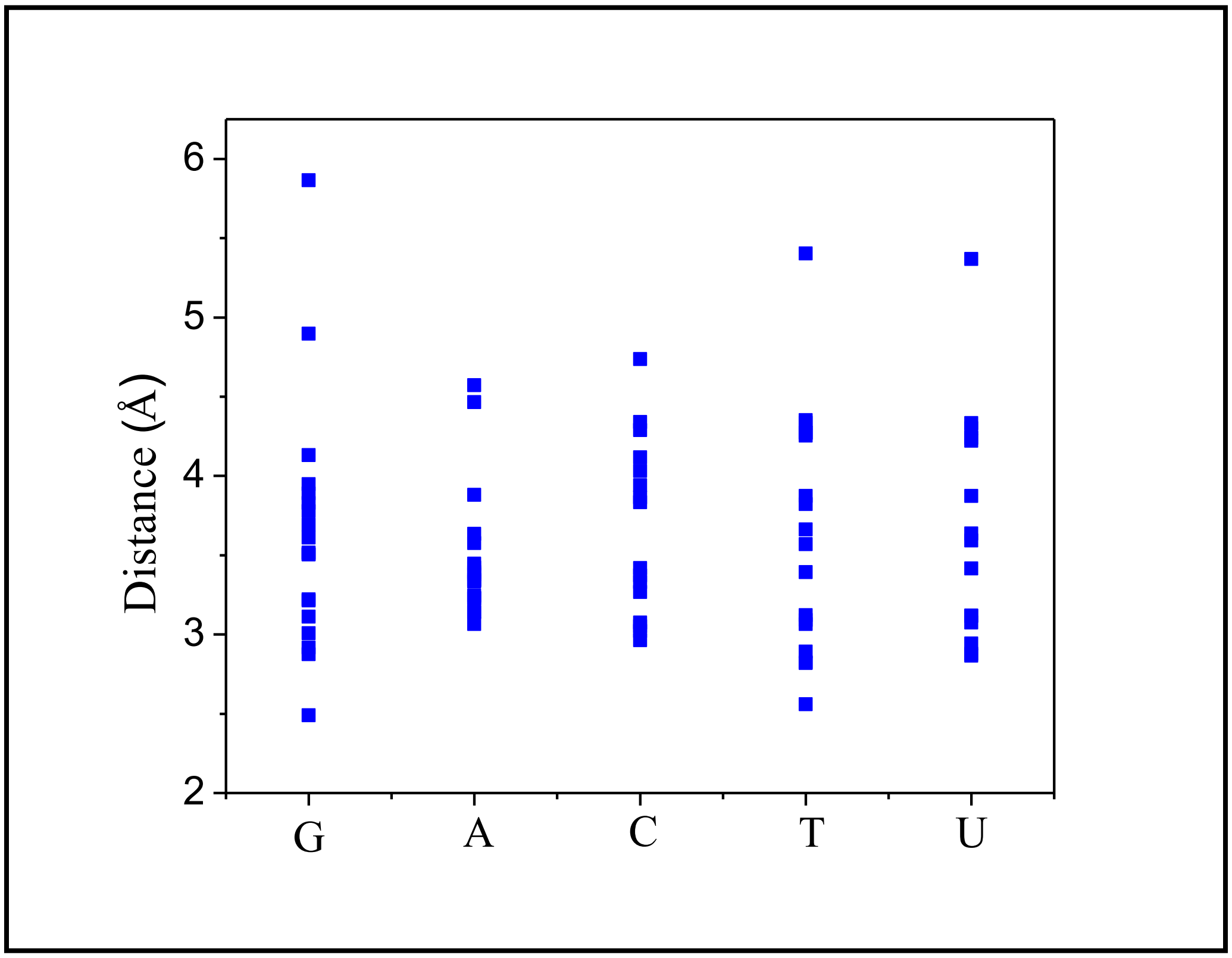

***Figure 3: The distance between the nucleobases atoms and the tubular surface atoms in the equilibrium configurations of BNNT conjugates.***

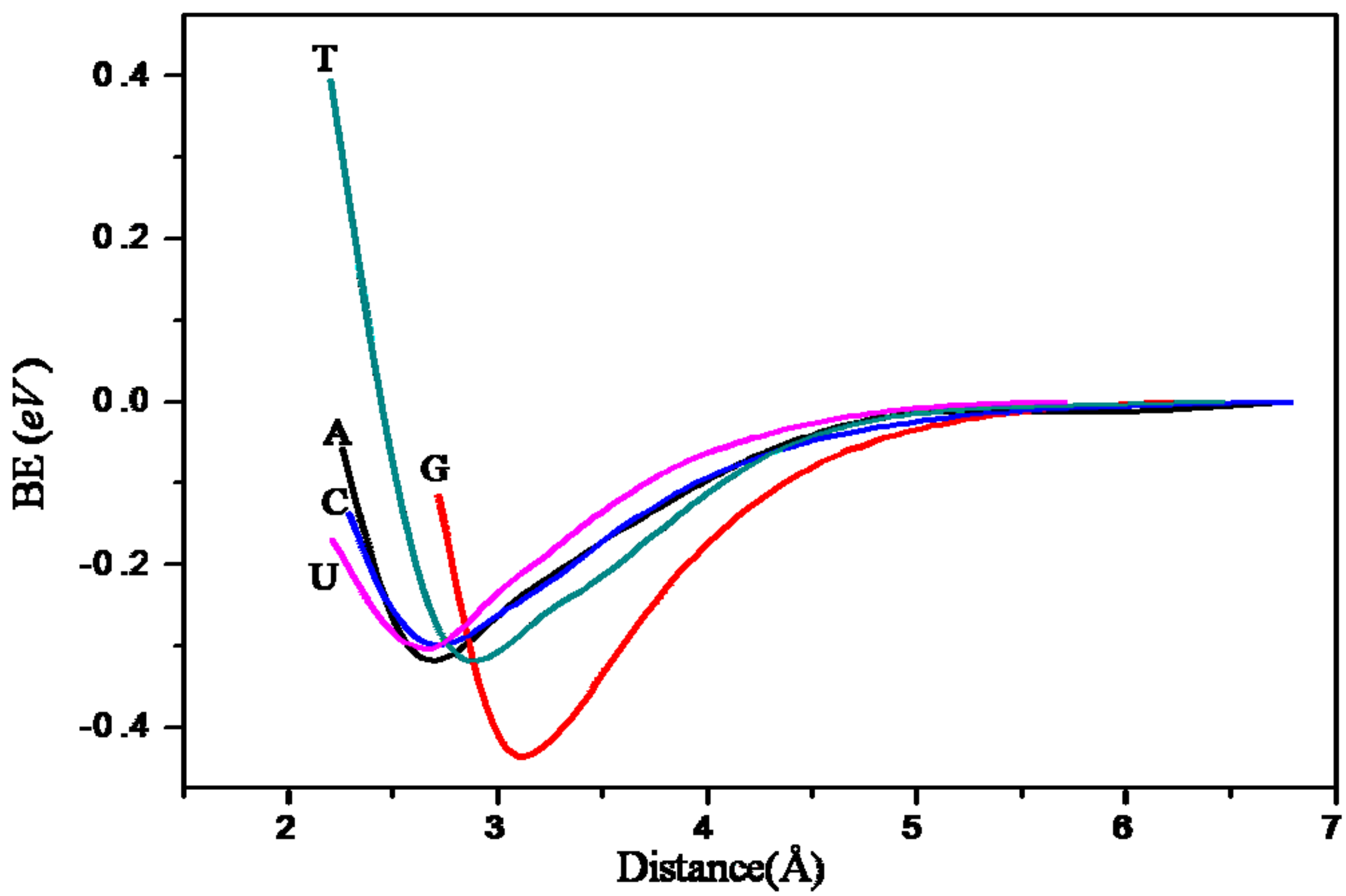

***Figure 4: (color online) The potential energy variation of the nucleobases interacting with BNNT as a function of the distance. The distance represents the separation between the center of the mass of the tubular surface and that of the base. A, G, C, T, U are represented by black, red, blue, green and pink lines, respectively. Zero of the energy is aligned to the non-interacting regime of the surface.***

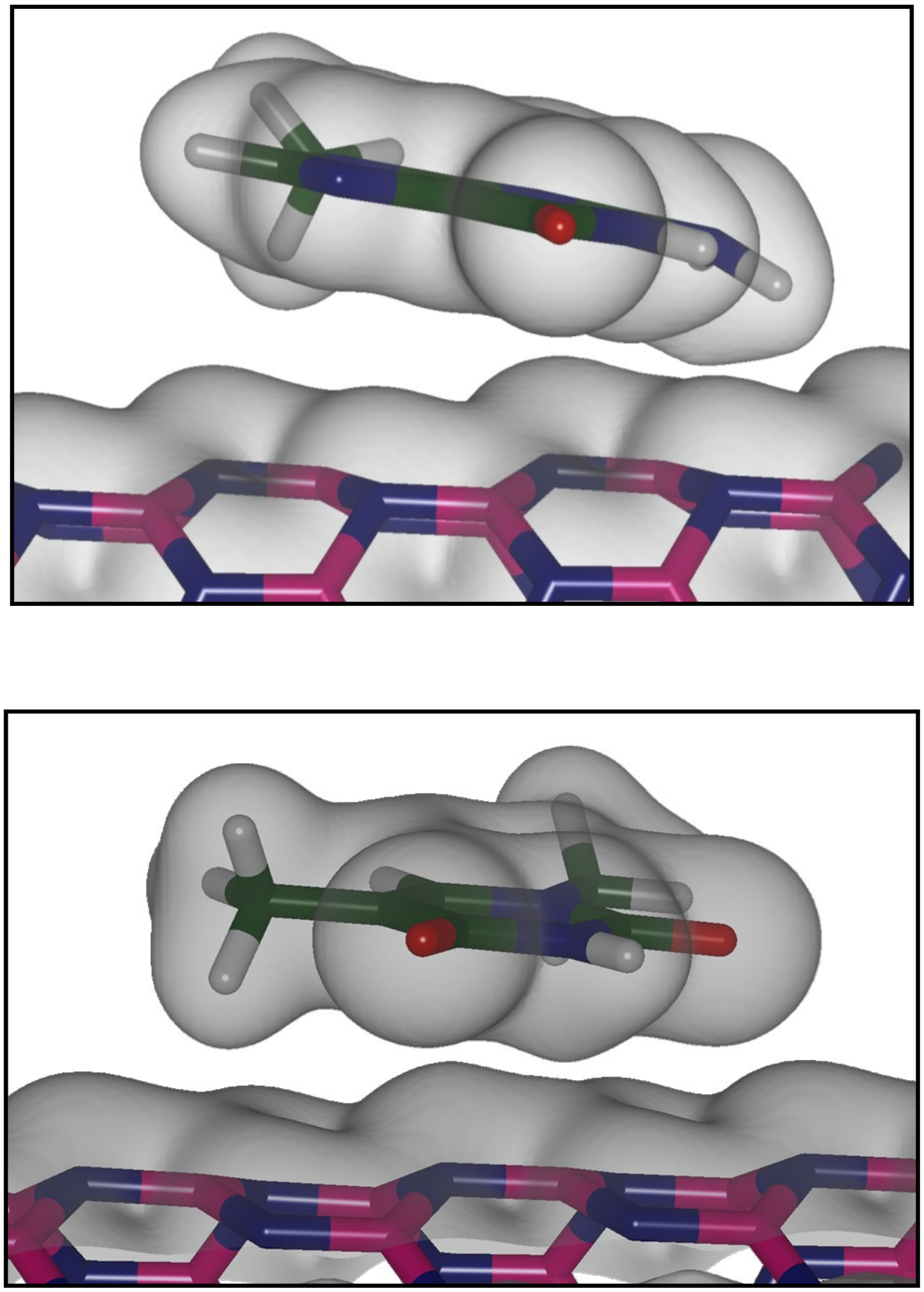

***Figure 5: Total charge density of guanine (above) and thymine (below) conjugated BNNT. The isosurface levels were set at 0.08 bohr $^{-3}$. The pink, blue, green, red and white sticks represent B, N, C, O and H, respectively.***

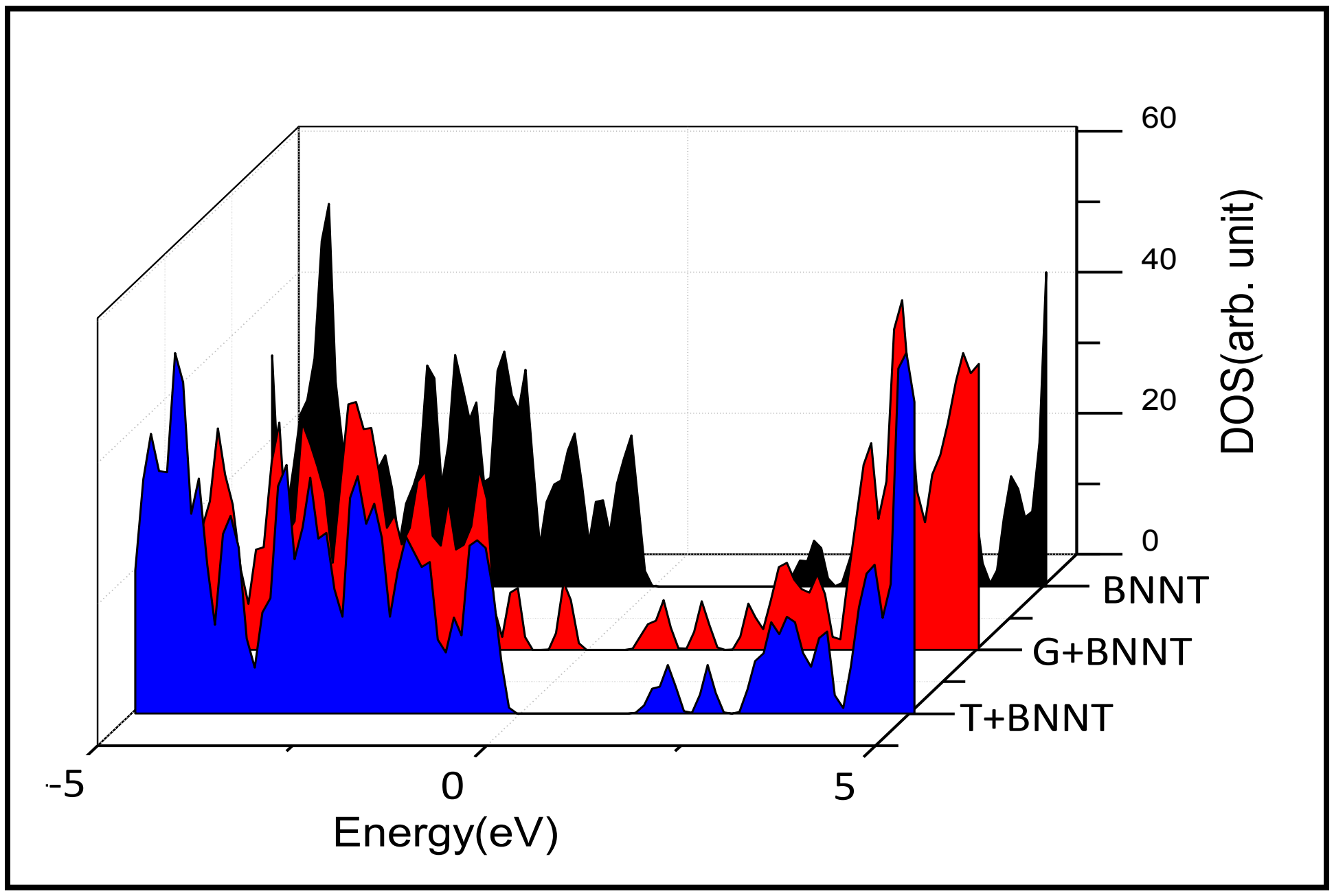

***Figure 6: Density of states of a pristine BNNT and guanine and thymine conjugated BNNT. Zero of the energy is aligned to the top of the valence band.***